\documentclass[fleqn,twoside,onecolumn,nofootinbib,showkeys,11pt]{revtex4-1} 
\usepackage{caption2}

\usepackage{subfigure}
\usepackage[utf8]{inputenc}
\usepackage[pdftex]{color,graphicx}%
\usepackage[bookmarks=false]{hyperref}
\usepackage{cleveref}%
\usepackage{float}

\begin{document}
\bibliographystyle{plainnat}
\UseRawInputEncoding 

\title[A method of estimating the content of crystalline phases in multiphase samples based on X-ray diffraction data]
{A METHOD OF ESTIMATING THE CONTENT OF CRYSTALLINE PHASES IN MULTIPHASE SAMPLES BASED ON X-RAY DIFFRACTION DATA}%

\author{S.V.\,Gabielkov}%1
\author{I.V.\,Zhyganiuk}%1
\author{A.D.\,Skorbun}%1

\affiliation{Institute for Safety Problems of Nuclear
Power Plants,   Nat. Acad. of Sci. of Ukraine, \\ 36a, Kirova str., Chornobyl 07270, Ukraine}%1
\email{i.zhyganiuk@ispnpp.kiev.ua}%e-mail 1

\begin{flushleft}
PACS 28.41.Kw
\end{flushleft}

\begin{abstract}
A method for estimating the relative content of crystalline phases of a multiphase sample, based on probabilistic analysis of the intensities of the diffraction pattern reflexes, has been developed. The method is based on the introduction of some numerical parameter, which uniquely characterizes the diffraction pattern, and comparison of the diffraction patterns according to this parameter, using methods of computational statistics. In situations where the absorption coefficients of the phases are close, the method works solely on the analysis of the diffraction pattern data from a specific multiphase sample.

It is shown on the control measurements that the uncertainty of the estimates does not exceed $\pm 8$~\%.
\end{abstract}

\keywords{powder diffraction, quantitative phase analysis, correlation, computational statistics}

\maketitle
\thispagestyle{empty}

\section{Introduction: problems of phase content estimation in X-ray phase analysis}

The procedure for quantitative analysis of X-ray phase analysis data of multiphase samples can be conditionally divided into two parts. First, the phases are identified in the diffraction pattern, and then the content of the detected phases is estimated based on the ratios of the reflex intensities. The generally accepted analysis procedure becomes difficult, if not impossible, in two extreme cases: when the number of detected reflexes becomes very large (hundreds of reflexes), and when the reflexes are difficult to distinguish from the noise background, which leads to the possibility of false identification of non-existent reflexes, or losing true ones. However, even in the case of two phases, the assessment of the phase content today requires a comparison of the intensity of the main reflexes of the experimental diffraction pattern with reference calibration curves.

The proposed work is the third in a series of methodical works by the authors, in which the problems of X-ray diffraction data analysis from samples of multiphase materials with low phase content are consistently considered. All of them are based on a probabilistic approach to determining certain parameters of diffraction patterns, which in turn is based on the obvious premise that in reality complex diffraction patterns are analyzed with a certain degree of reliability, especially in cases of noisy diffraction patterns with a low phases content~\citep{8Gabielkov}, when reflexes overlap each other, hide in the noise, etc. All this gives reason to consider such cases as a situation when in a real diffraction pattern, a tabular diffraction pattern of a phase, known from databases, is realized only with a certain probability.

In the paper~\citep{Skorbun}, the problem of identifying the reflexes in the diffraction pattern was considered for the case when the intensities of these reflexes are at the level of noise. This problem was solved by using correlation analysis: finding a correlation between the selected part of the diffraction pattern and the test line, the parameters of which are close to the parameters of the reflexes of the experimental diffraction pattern. The key in this approach is the calculation of the degree of correlation, based on the methods of computational statistics~\citep{Moore, Simon, PanasyukSkorbun}. This method made it possible to increase several times the possibility of determining the presence of reflexes for a given angle and, accordingly, to increase the reliability of estimating the intensity of reflexes whose amplitude is at the noise level.

In the work~\citep{skorbun2021method}, a probabilistic approach to the identification of not individual reflexes, but individual phases, which in such a problem are considered as a set of intensity values for given angles, was proposed. This problem was solved by searching for correlations between two samples, one of which is an experimental diffraction pattern, and the other is a diffraction pattern of some compound, obtained from the database, the presence or absence of which in this sample must be established. The degree of correlation is also calculated based on methods of computational statistics~\citep{Simon, Moore}. The correlation method developed in~\citep{skorbun2021method} in the case of noisy diffraction patterns turned out to be much more effective than the usual procedure for diffraction pattern identification, which consists in a "manual" comparison of the obtained experimental diffraction pattern with information from databases. Therefore, the second task, namely the phase identification task, especially phases with a low content, in diffraction patterns with a large number of phases can also be considered as resolved.   

However, the task of quantitative determination of the content of phases in the above conditions of low content and noisy diffraction patterns remains problematic. The usual analysis procedure for determining the content of phases consists in comparing the intensities of the most intense lines of different phases and using standards~\citep{MudryiKulik,Madsen}. With a large number of low-intensity reflexes from low-content phases and the absence of standards, this procedure is not effective due to the high probability of incorrect identification of weak reflexes.

The basis of a given work are the results obtained in~\citep{ Moore, PanasyukSkorbun, Skorbun}. In this work, the phase identification method based on automated correlation analysis~\citep{skorbun2021method}, including a known software for phase identification, is extended to the task of quantitative assessment of the content of individual phases.

All diffraction patterns in the databases are normalized so, that the intensity of the main reflex is equal to one thousand. Therefore, the intensities of the diffraction pattern reflexes, obtained in the experiment, are relative values. As a result, changes in the experimental intensities of the reflexes cannot be compared in reality with those in the table, and calibration curves obtained from the analysis of reference mixtures are used for quantitative evaluations.

The idea of the presented work is based on the fact, that for each compound, that is, for its corresponding diffraction pattern, it is possible to introduce some characteristic parameter by which the intensities of these diffraction patterns can be compared with each other. As such parameter it was proposed to use the sum of reflexes intensities of the normalized tabular diffraction pattern. In combination with the fact that the value of the intensities is tied to certain angles, this parameter is an unambiguous characteristic of the diffraction pattern of the given compound.

\section{Diffraction patterns for analysis}

In this work, diffraction patterns from databases were used to develop and demonstrate the capabilities of the method. The authors' initial attempts to develop a method based on their own experimental diffraction patterns ran into the following problem. To demonstrate the capabilities of the method, it must be demonstrated on fully and reliably decoded diffraction patterns in order to know in advance the correct answer, with which the result of the developed method will be compared. Such a result can be obtained, for example, using standards. But when creating standards, the problem of the reliability of the created benchmarks arises. And in the final result, experimentally obtained diffraction patterns of standards, are again analyzed by comparison with diffraction patterns from databases. Therefore, in this work, diffraction patterns from multiphase samples for analyzing by the developed method were created as a simple combination of diffraction patterns taken from databases. And reference diffraction patterns then have been used to demonstrate the efficiency of the method. This means that we are considering a situation, where various corrections, such as self-absorption, are insignificant.

\section{Correlation analysis based on Monte Carlo methods}

The task of determining the presence and quantitative content of certain crystalline phases in X-ray powder diffraction methods can be reduced to the task of comparing two samples: an experimental diffraction pattern and a diffraction pattern from the database for the compound, whose presence is checked. From the point of view of statistics, the diffraction pattern represents two columns of numbers: a column of regular angles and a column of corresponding reflexes intensities. If the reflexes intensities of different diffraction patterns are compared for the same angles, the columns of the angles can be disregarded further, and then the problem of the diffraction pattern decoding can be reformulated as the problem of comparing the intensities for the corresponding angles, and, therefore, as the problem of comparing paired samples of numbers which, due to uncertainties of estimates of reflex intensities, due to the possible overlapping of reflexes and even the possible absence of some reflexes, can be considered as a task of comparing two random (in the specified sense) paired samples. And, therefore, as a task for mathematical statistics.

Calculations are based on the use of methods of computational statistics~\citep{Moore} which allow to determine the presence of correlation between samples with increased reliability. This way of correlations estimation by Monte Carlo methods was modified for the possibility of calculating correlations between well-correlated samples~\citep{PanasyukSkorbun}, which, as it turned out, made it possible not only to calculate the degree of correlation between two samples, but also to obtain a fundamentally new result: the result of correlations calculation turned out to be sensitive to the absolute values of sample elements, i.e. to the line intensities.  In~\citep{Skorbun}, the method was extended to the analysis of diffraction patterns obtained in X-ray phase analysis to search for weak lines, and in~\citep{skorbun2021method} it was successfully applied to the identification of phases in complex diffraction patterns based on the search for correlations between test diffraction patterns from databases and experimental ones.

In the methods of computational statistics~\citep{Simon, Moore}, based on the use of Monte Carlo methods, the correlation between two samples is calculated not as an analogue of the Pearson coefficient in the form of a number from zero to one, but as the probability of accepting the hypothesis of the presence or absence a correlation between them, based on the analysis of dispersion of the value of the numerical indicator of the degree of correlation, obtained by the methods of computational statistics. When using the modified method of computational statistics~\citep{PanasyukSkorbun, Skorbun}, the correlation value is calculated in relative units, because it depends on the parameters of the samples.

\section{Calculation}

If (at the initial stage of the analysis) one does not pay attention to the need to take into account various corrections, then at first glance it may seem that the ratio of the intensities of the maximal reflexes can serve as the necessary parameter for comparing two different phases. Next, the necessary corrections obtained, for example, from the measurement of reference mixtures, are made in this relation. However, precisely because of the large number of factors that can affect this ratio, the existing practice of determining the phase ratio faces great difficulties, and the analysis of these factors is practically a mandatory procedure for preparing diffraction patterns (see, for example, the Rietveld method~\citep{Toby}. In practice, such works require the creation of standards, which is far from always possible. There is also the question of the uncertainty of the result of such an assessment. 

 Therefore, more unambiguous characteristics of the diffraction pattern should be used as a parameter that characterizes the phase. For example, as in this work, use the correlation characteristic of the sample described below (the diffraction pattern is considered as a sample of numbers ordered by angles). The following is meant. Information (angles or interplanar distances, and corresponding reflexes intensities) about all possible phases is already in the databases (crystallographic tasks are not considered at the moment). Based on the approaches of computational statistics, it is possible to introduce some correlation characteristic that uniquely characterizes the diffraction pattern.

\section{The parameter of mutual correlation of the diffraction patterns}

Let's consider tabular diffraction patterns with intensity normalized to 1000. Let us introduce the notation: the number of regular angles for which a diffraction pattern was obtained, and therefore the size of the sample, will be denoted by $N$. The numbers (signal intensities for each angle) in the experimental sample will be denoted as~$g_{i}$ and $i=1, 2, \ldots , N$, and the values of the intensities in test diffraction pattern from the database, respectively, through $b^{i}$. In computational statistics~\citep{Moore} the degree of correlation between two samples is proportional to $S = \sum^{N}_{i} g_{i} b_{i}$, and therefore, depends on the values of the elements of the samples, will be different for different diffraction patterns, and therefore is given in the following in relative units. We emphasize once again that we are interested not just in the value of $S$, but in finding the most probable value of $S$ when analyzing our samples, taking into account possible uncertainties (errors) in the values of both $g_{i}$ and $b_{i}$.

\begin{figure}[H]
\vskip1mm
{\includegraphics[width=0.47\textwidth]{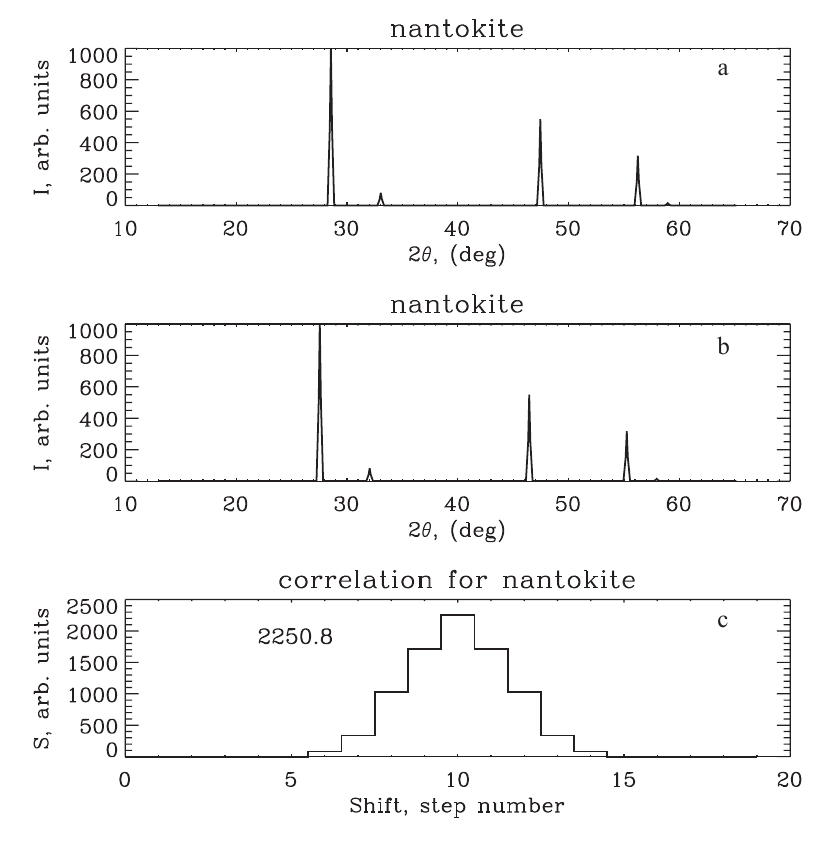}}\hfill
{\includegraphics[width=0.47\textwidth]{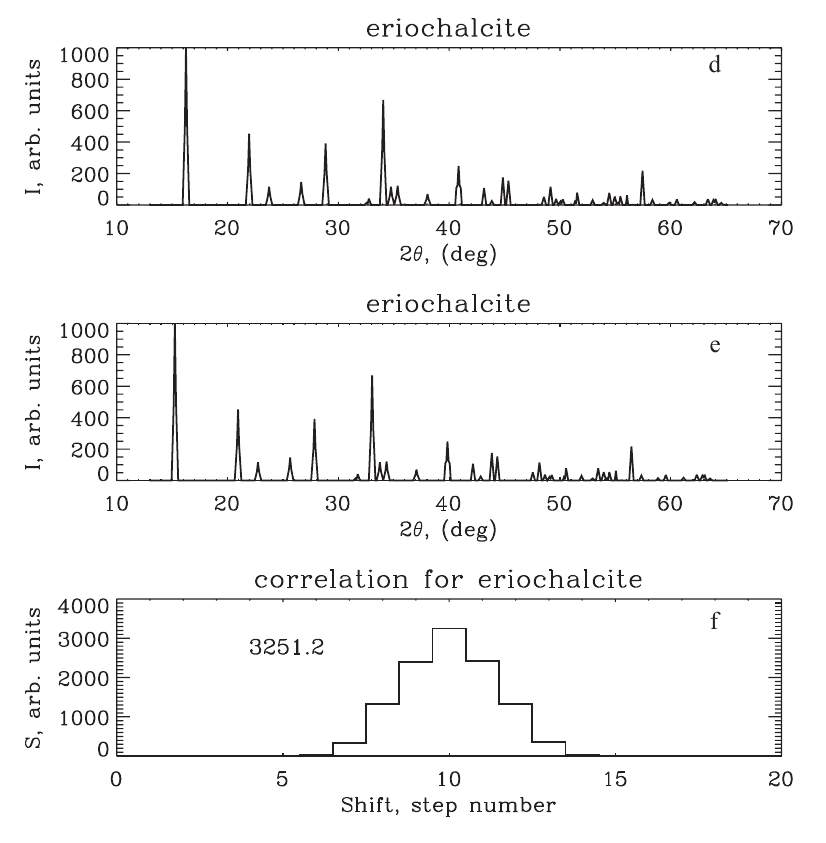}}\hfill
\caption{The result of scanning the ''experimental'' diffraction patterns of nantocite and eriochalcite with the same diffraction patterns, the intensity of which is normalized to 1000. Left: nantocite. Right: eriochalcite. The resulting match degree histograms have the expected peak at step 10. The value of $S_{max}$ is shown in the figure.
}\label{f1}
\end{figure}

It is easily verified (see also~\citep{Skorbun, skorbun2021method}) that the value of $S$ will be maximal when comparing identical, that is, fully correlated samples. The degree of self-correlation $S_{max}$ for a given sample (phase) will be different for different phases. And although in the proposed approach, the result of assessing the degree of correlation in such calculations is obtained in relative units, however, for some specific calculation algorithm, the result will be unambiguous and can be compared for different phases. This point is illustrated in Figure~\ref{f1} for diffraction patterns of nantokite and eriochalcite taken from the Crystallography Open Database (COD)~\citep{Cedric}. The diffraction patterns are taken from the database, where it is customary to normalize the intensity of the maximum reflex by 1000.

Graphs in Figure~\ref{f1} are obtained in the following way. Figure~\ref{f1}~a shows a modelled ''experimental'' diffraction pattern that needs to be identified. The test diffraction pattern in Figure~\ref{f1}~b (that is, the same diffraction pattern as in Figure~\ref{f1}~a) by 10 elements-angles of the sample is shifted to the left (not shown to scale) relative to the experimental one. Next, we scan the test diffraction pattern along the ''experimental'' one, shifting by one step to the right and calculating $S$ at each step. With the best match between the samples, we will get the maximum value of $S = S_{max}$ for this specific diffraction pattern. Since two identical diffraction patterns (samples) are compared, the maximum value of the degree of correlation will be obtained at given step~$10$. This procedure takes into account the possible variability of the angles for the corresponding reflexes in practice. In Figure~\ref{f1}~c it is seen that the degree of correlation when comparing identical samples is $S_{max} = 2250.8$ for nantokite. A similar calculation for eriochalcite gives $S_{max} = 3251.2$. That is, $S_{max}$ for different compounds is noticeably different. The designation ''max'' means that with the accepted method of normalizing the intensities of diffraction patterns by $1000$, $S_{max}$ is the maximum possible value of the degree of mutual correlation for a given sample (diffraction pattern), that is, when it is correlated with itself.

%Ðèñ.1

\begin{figure}[H]
\vskip1mm
\begin{center}
\includegraphics[width=0.74\textwidth]{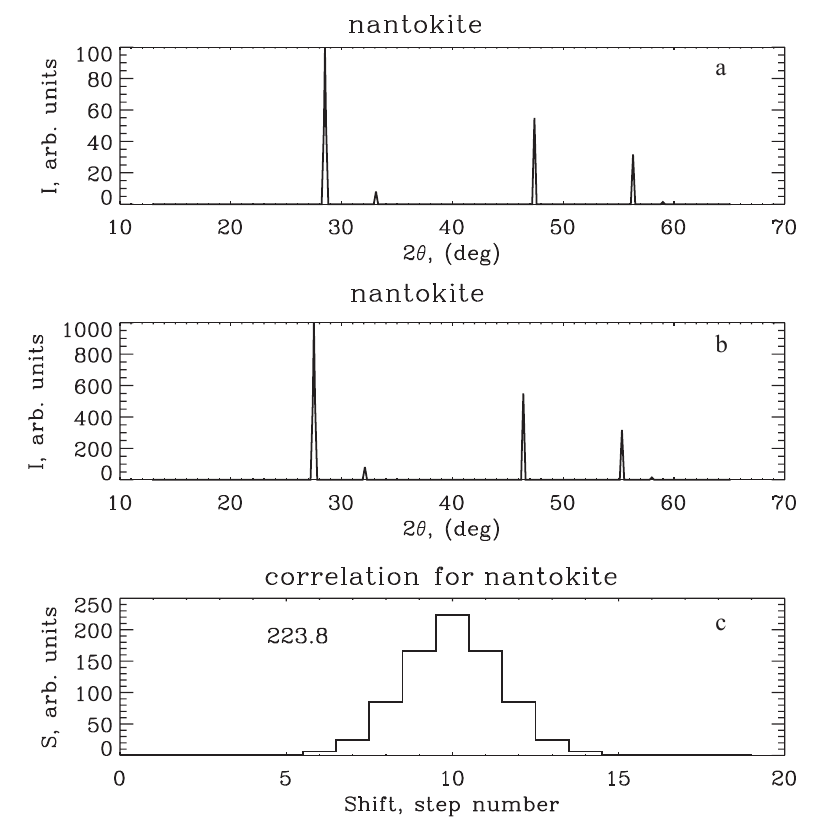}
\end{center}
\vskip-1mm
\caption{The result of scanning the model diffraction pattern of nantocite, the intensity of which is $10$ times lower, with the reference diffraction pattern of nantocite. Accordingly, the value of $S$ is also decreased.
}\label{f2}
\end{figure}

Please note that although in the table data, the intensities of the main reflexes for different compounds are the same $(= 1000)$, the values of $S_{max}$ are different. This fact is the basis for further calculations. Any changes in the intensities of the lines of the experimental diffraction pattern of some compound will result in a decrease in the value of $S$ compared to $S_{max}$. The demonstration of this is shown in Figure~\ref{f2}, which shows the result of calculating the degree of correlation $S$ between two samples for the case when both samples are the same in structure (diffraction pattern of nantokite from the COD 96-901-3925 database~\citep{Cedric}), but have different intensities. Figure~\ref{f2}~a shows the diffraction pattern of nantokite, which we take as the experimental one. The intensity of the reflexes in it, is ten times less than the intensity of the reflexes of the same diffraction pattern, normalized by 1000, which is shown in Figure~\ref{f2}~b and which we consider as the test one. The result of the correlation calculation is shown in Figure~\ref{f2}~c. The maximum value of Smax for the test calculation for nantokite, as indicated above, is equal to~$2239$. In the case shown in Figure~\ref{f2}, when the intensity of all lines in the experimental diffraction pattern is reduced tenfold, the value of $S$, that is, the degree of correlation between these diffraction pattern, also decreases tenfold and is equal to~$223.8$. This result, i.e. the resulting decrease in $S$ compared to $S_{max}$ opens the way to estimating the relative content of some crystalline phase in multiphase materials.

%Ðèñ.2

\section{Algorithm for phase content estimation}

\begin{figure}[H]
\vskip1mm
\begin{center}
\includegraphics[width=0.44\textwidth]{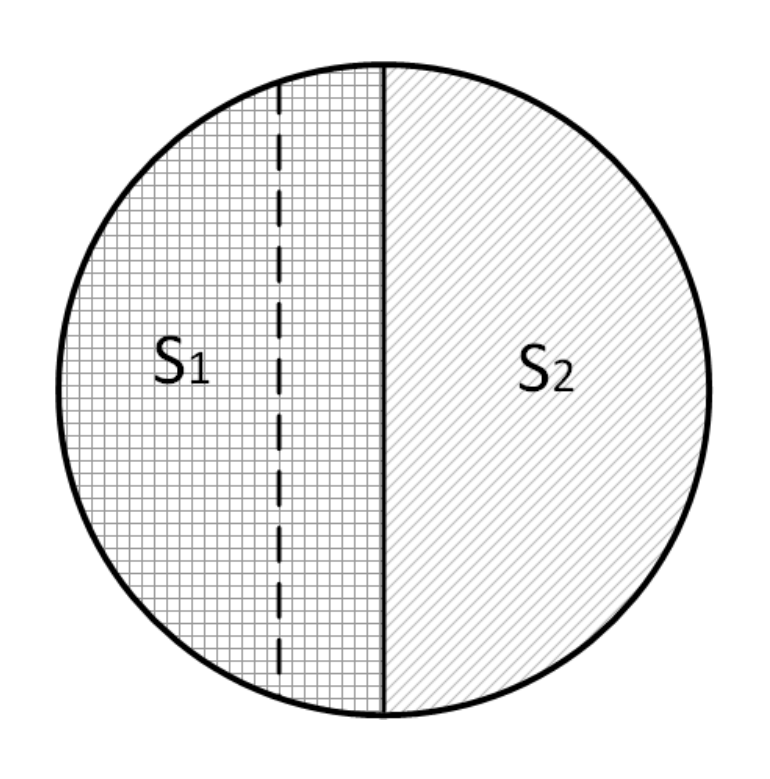}
\end{center}
\vskip-1mm
\caption{Scheme of measurements of the model standard. Dashed line - division of the cuvette in the ratio 1:2.
}\label{f3}
\end{figure}

First, we formulate the conditions of the issue and the assumptions under which the calculations are carried out. Let's imagine the measurement scheme (Figure~\ref{f3}), when the sample (cuvette with powder) is divided into two equal parts. Each of them is a separate phase. This is an approximation of the classical situation of the analysis of a two-phase system with the same absorption coefficients. In such an experimental scheme, there is no mutual influence of phases. Accordingly, the total diffraction pattern will be the sum of the diffraction patterns from individual phases in an equal ratio of their content.

\begin{description}
	\item[1] The paper considers the possibility of comparing only the relative contents of the identified phases. That is, the presence of signals hidden in noise or the contribution of the amorphous phase are not taken into account.
	\item[2] If the phases are in equal ratios, then under all conditions the ratio of reflexes intensities or squares should reflect this ratio.
	\item[3] Corrections such as self-absorption, texture effects, etc. are not considered at this stage, since they are not taken from the results of a specific measurement (a specific diffraction pattern), but from other considerations (from the analysis of calibration curves, theoretical estimates, etc.).
	\item[4] The total relative content of all identified phases must be equal to unity.
	\item[5]  The signal/noise ratio must be taken into account: in theory, the signal can be as small as desired, but in reality, there is a limit: the minimum permissible signal/noise must be greater than one.
	\item[6] It is believed~\citep{MudryiKulik, Madsen} that the intensity of reflexes from this phase is ideally proportional to the content of this phase. If there are two phases with a content of 50~\% each (Figure~\ref{f3}), then it is obvious that the ratio of experimental intensities should reflect this 50~\% ratio of the content of phases. Mixing the sample, if the absorption coefficients are equal, does not change the situation.
	\item[7] In the paper, it is proposed to use such a value as the sum of the intensities of all lines of their standard diffraction pattern 
$S _{\rm 1,2}^{\rm (0)}  = \sum\limits_i^n { I_i }$ to characterize individual diffraction pattern, where n is the number of reflexes of the standard diffraction pattern. It is obvious that this value, like the intensity of the main reflex, is proportional to the phase content. If, for example, in Figure~\ref{f3} to move the dividing line so that the area ratio becomes 1:2 instead of 1:1 (dotted line), this should be clearly reflected in the ratio of $S_{expr}$ values for the phases of the sample in question: $ S_{1} / S_{2} = const$ should change to $ S_{1} / S_{2} = const^{*} (1/2) $.
\end{description}

%Ðèñ.3

\section{Formulas for calculating the content of complex model diffraction pattern}

It was demonstrated above (Figure~\ref{f2}) that the proposed method allows to distinguish reference samples for the same phases with different contents. Now let's check the possibilities of the method on model mixtures of phases. Let's take the tabular diffraction pattern of nantokite and eriochalcite, which are normalized to 1000. Let's create a model diffraction pattern of a mixture of these phases, combining them into one. We will remind that the problem of identification of individual phases in such mixtures was solved by statistical methods in~\citep{skorbun2021method}. Now we are interested in the possibility of quantitative determination of the relative content of these two phases, of which it is consisted, which we are going to estimate by the value of $S$.

 %Ðèñ.4
\begin{figure}[H]
\vskip1mm
\renewcommand{\thesubfigure}{\arabic{subfigure}}
{\includegraphics[width=0.47\textwidth]{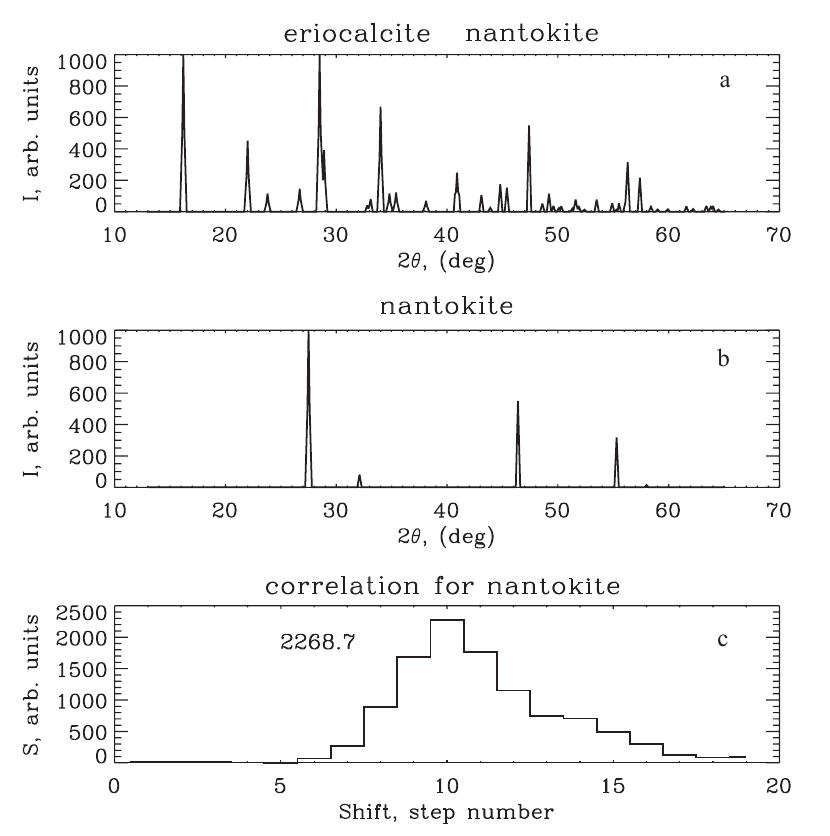}}\hfill
{\includegraphics[width=0.47\textwidth]{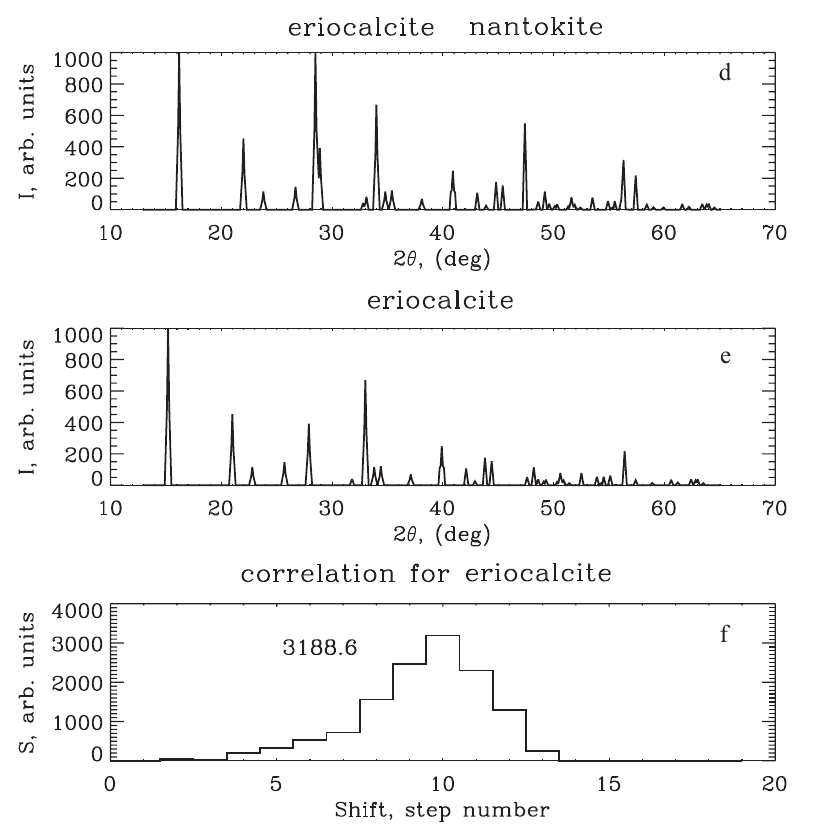}}\hfill
\caption{The result of the analysis of the diffraction pattern of the model mixture of two compounds (see text) in the ratio of 50--50~\%. On the left --- a search for the nantokite phase, on the right --- a search for eriochalcite. a): model (table) diffraction pattern of the mixture; b): diffraction pattern of the phase that is being sought; c): histogram of the degree of correlation of the experimental phase with the tabular one.
}\label{f4}
\end{figure}

Let's consider the model situation of a mixture of two phases in equal concentrations, when the intensities of the most intense reflexes of each phase are normalized by 1000, as shown in Figure~\ref{f1}. The upper graphs of Figure~\ref{f4} shows the total diffraction pattern of the mixture of these two phases, in the middle is the tabular diffraction pattern of the phase that is searched for in the mixture (nantokite or eriochalcite), and below is the result of the searching for the corresponding phase. Since the maximum reflexes of both phases are the same and equal to 1000, the $S_{expr } / S_{max }$ values for each phase will be equal to one (see Figure~\ref{f1}). In this case, the total value of relative contents for two ($k = 2$)  phases  
$S_{calc}  = \sum\limits_i^k {\frac{S_{expr}^{(i)}}{{S_{max }^{(i)} }}}$. But the model assumes that these two phases are in equal proportions, the content of each phase is  $1/2$. Therefore, the correct ratio for individual phases is obtained as 
$(S_{expr}^{(i)} /S_{max }^{(i)} )/S_{calc}$.

Therefore, with arbitrary (and unknown) ratios between phases
$\sum\limits_i^k {\frac{{S_{expr}^{(i)} }}{{S_{max}^{(i)} }}}  = S_{calc}$, and for each phase its contribution is 
	\begin{equation}\label{eq:1}
		(S_{expr}^{(i)} /S_{max}^{(i)} )/S_{calc}
	\end{equation}

An example of such analysis for model diffraction patterns of a mixture of two phases with the same content (Figure~\ref{f4}) is given in Table~\ref{tab1}.

\begin{table}[H]
\noindent\caption{Results of finding the phase ratio in model samples.}\vskip3mm\tabcolsep4.5pt
\begin{center}
{
\begin{tabular}{|l|c|c|c|c|c|}
 \hline%
   \multicolumn{1}{|l}{Phase}%
 & \multicolumn{1}{|c}{$S_{max}$}
 & \multicolumn{1}{|c}{$S_{expr}$}
 & \multicolumn{1}{|c}{$S_{expr} / S_{\max}$}
 & \multicolumn{1}{|c}{$S_{calc}$}
 & \multicolumn{1}{|c|}{$(S_{expr} / S_{max}) / S_{calc}$}\\
\hline%
 Nantokite & $2250.8$ & $2268.7$ & $1.008$ & $1.989$ &  $0.5068 \,\, (50.68 \%)$\\ 
 Eriochalcite & $3251.2$  & $3188.6$ & $0.981$ & $1.989$ & $0.4932 \,\, (49.32 \%)$ \\%
\hline
\end{tabular}}\label{tab1} 
\end{center}
\end{table}

It can be seen from the last column that the accuracy of the calculations is no worse than $\pm 0.5~\%$ (the phase ratio of $50~\% : 50~\%$ was included in the model).

Therefore, the $S_{exp} / S_{max}$ values give the relative value of the content for the two selected phases. A sample may consist of many identifiable phases as well as unidentified and/or amorphous phases. Their total content is given by the ratio
\begin{equation}\label{eq:2}
		S=\sum\limits_i^N {\frac{{S_{expr}^{(i)} }}{{S_{max}^{(i)} }}}  + B  \equiv 1,
\end{equation}
where B is the contribution of unidentified reflexes, haloes, etc.

However, the performed analysis is correct for any pair of phases and does not depend on the content of the others. This can be easily understood by imagining that the image of the phases in Figure~\ref{f2} is only part of a larger sample. The value B is estimated from other reasons, outside the scope of this approach. Within the framework of the developed method, the value of B is not taken into account, but only the ratio of the content of the identified phases is evaluated.

If the number of identified phases is greater than two, the procedure for analyzing the general diffraction pattern does not change: the ratio of the contents of any two phases is calculated according to the described algorithm.

\section{Method testing}

\begin{figure}[H]
\vskip1mm
\renewcommand{\thesubfigure}{\arabic{subfigure}}
{\includegraphics[width=0.47\textwidth]{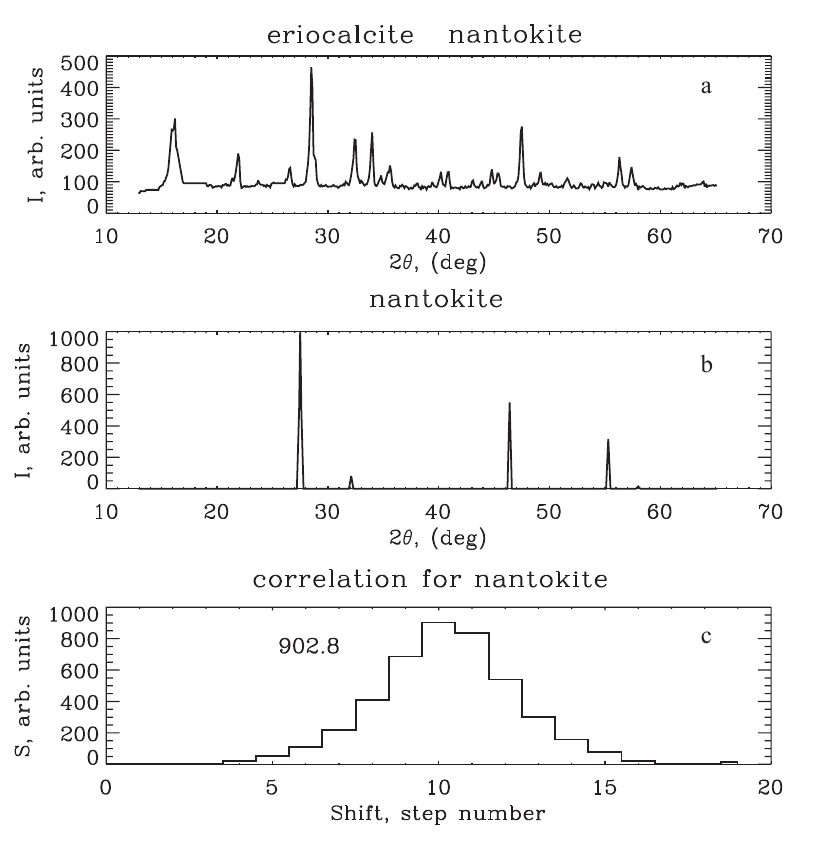}}\hfill
{\includegraphics[width=0.47\textwidth]{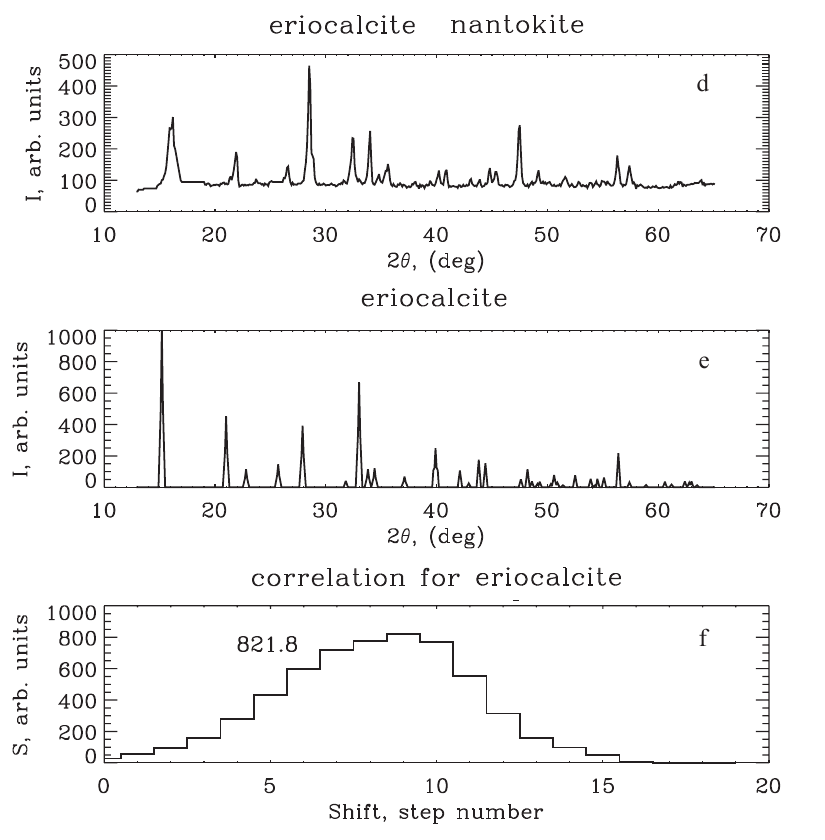}}\hfill
\caption{The result of the analysis of the phase content of the diffraction pattern of a mixture of two compounds (see text) in the ratio of 70--30~\%. On the left --- a search for the nantokite phase, on the right --- a search for eriochalcite one. a): experimental diffraction pattern of the mixture; b): diffraction pattern of the phase that is being searched for; c): histogram of the degree of correlation of the experimental phase with the tabular one.
}\label{f5}
\end{figure}

As already it was mentioned, the workability of the approach to the problem of phase identification by calculating correlations, based on Monte Carlo methods, is substantiated and demonstrated in~\citep{Skorbun, skorbun2021method}. It is shown above that the method works in the case of ''ideal'' tabular diffraction patterns. It remains to show the possibilities of this method for estimating the content of phases in multiphase materials on real data. Since it is believed that this can be done only by having a calibration curve obtained on the basis of standards, the developed method of estimating the phase content must be verified by analyzing the diffraction patterns of the calibrated mixtures. For this purpose, a reference mixture of nantokite and eriochalcite in the proportions of $70~\% : 30~\%$ was used. The graphical results of the analysis are shown in Figure~\ref{f5}. Calculations for basic diffraction patterns for estimating $S_{max}$ see in Figure~\ref{f1}. The results of calculations are summarized in Table~\ref{tab2}. The obtained result: the content of nattocite is $60~\%$, and the content of eriochalcite is $40~\%$ (instead of the expected $70~\%:30~\%$). The discrepancy can be explained either by the inaccuracy of the preparation of the mixtures, or by the fact that this is the real uncertainty of the method with unknown peculiarities of the experiment. Most likely, the final answer requires an analysis of metrological standards.

 %Ðèñ.5

\begin{table}[H]
\noindent\caption{Results of searching for phase content ratios in experimental samples.}\vskip3mm\tabcolsep4.5pt
\begin{center}
{
\begin{tabular}{|l|c|c|c|c|c|}
 \hline%
   \multicolumn{1}{|l}{Phase $(70~\% : 30~\%)$}%
 & \multicolumn{1}{|c}{$S_{max}$}
 & \multicolumn{1}{|c}{$S_{expr}$}
 & \multicolumn{1}{|c}{$S_{expr} / S_{\max}$}
 & \multicolumn{1}{|c}{$S_{calc}$}
 & \multicolumn{1}{|c|}{$(S_{expr} / S_{max}) / S_{calc}$}\\
\hline%
 Nantokite    & $2250.8$  & $902.8$ & $0.401$ & $0.65$ & $0.617  \,\, (61.7 \%)$\\ 
 Eriochalcite & $3251.2$  & $821.8$ & $0.253$ & $0.65$ & $0.389  \,\, (38.9 \%)$ \\%
\hline
\end{tabular}
}\label{tab2} 
\end{center}
\end{table}
 
\section{Estimation of the uncertainty of the method}

Let's evaluate the uncertainty of the calculation algorithm and, accordingly, the method itself. The need for such an estimate stems from the fact that in reality for a specific diffraction pattern it is only possible, based on previous experience, to attribute the accuracy with which the intensity of the main line is determined. As for weak lines, it is almost impossible for them to reasonably estimate the uncertainty of their intensity on the basis of a separate diffraction pattern.

The uncertainty of the result of statistical calculations can be seen from the Table~\ref{tab1}: the difference between the result and the result embedded in the model is approximately $0.5~\%$.

Let us assume that the uncertainty of the estimation of the intensity of all reflexes is the same and is arbitrarily chosen equal to $\pm 5~\%$. That is, the diffraction pattern is a sample of numbers whose experimental values are within $\pm 5~\%$ around some unknown ''exact'' value. In this situation, it is proposed to estimate the range of possible values of $S_{exp}$, which is changed when the reflexes intensities are changed, as a measure of the uncertainty of the phase content assessment result. This can be done based on the use of Monte Carlo methods. Namely, the value $\pm \Delta$  from the range $\pm 5~\%$ is randomly added to each line of the tabular diffraction pattern. For such a modified sample, $S_{exp}$ is calculated. This procedure is repeated several thousand times. Next, you can build a histogram of the distribution of the obtained $S_{exp}$ values and take the most probable value as the correct result. The scatter around it (the width of the histogram) gives the uncertainty of the obtained estimate. It turned out that its width at half height is also approximately $\pm 5~\%$, but it should be noted that this is already the maximum value, that takes into account the variability of all lines of the diffraction pattern.

Uncertainty due to such effects as the presence of absorption, texture and other experimental factors does not depend on the described method and can be taken from the literature.

\paragraph*{\textbf{Warning}} The demonstrated efficiency of the proposed method of analysis of diffraction pattern, based on the above algorithm should not create the illusion of one hundred percent reliability of the obtained result. The development of the described method of correlation analysis was stimulated by the problems of identification and assessment of the content of phases in samples with their small content. In this case, the relative errors of determining the line intensities are increased sharply, up to the possible false identification of phases. Accordingly, the probability of erroneous determination of the content of this phase increases due to the fact that the uncertainty of the estimation of $S_{exp}$ is determined by the uncertainty of the estimation of the intensity of the corresponding reflexes. Since content estimates must necessarily be accompanied by estimates of their uncertainties, we recommend, as a mandatory element of the algorithm according to point 1, before starting calculations, to check the possible contribution of influence factors well known to practitioners (superposition of lines, self-absorption, etc.) and when describing the results of content estimation of individual phases, it is mandatory, as required by~\citep{Madsen, Guide}, to indicate by the direct text, the contribution of which factors were checked.

Of extreme importance in this method is the effect of distortion of reflexes intensities due to the superimposition of reflexes from different phases, the contribution of which can far exceed the accepted 5~\%. Therefore, it is recommended to carry out calculations through gradual iterations. First, determine the presence of one or another phase; then check that the ratio of intensities (especially sufficiently intense lines) corresponds to the basic diffraction pattern. If necessary, it is necessary to make appropriate corrections (by subtracting the contributions of reflexes of other phases), repeat the calculations after making corrections, and only then make a decision about the correctness of the $S_{exp}$ estimate.

\section{Conclusions}

\begin{enumerate}
	\item A method of estimation the relative content of crystalline phases in complex diffraction pattern of X-ray phase analysis is proposed. The idea of the method is that it is possible to enter a parameter, that uniquely characterizes the diffraction pattern.
	\item The task of identifying phases and estimating their relative content can be reduced to the task of probabilistic analysis of reflex intensities.
	\item The method is based on calculations of the degree of correlation between the experimental diffraction pattern and the diffraction pattern of the searched phase from databases. A parameter, based on the calculation of paired products of diffraction pattern, is introduced, by which they can be compared. The fundamental point is that the calculations are performed using computational statistics (permutation test).
	\item Calculations are made on the basis of some assumptions, for example, about the absence of the contribution of the absorption effect, and therefore the result of calculations by the proposed method requires a careful analysis of the correctness of the approximations. The error of statistical calculations is $\pm 0.5\%$. But the correctness of the assessment of the relative content of individual phases depends on the correctness of the assessment of the intensities of the diffraction pattern lines, making the necessary corrections in specific cases, etc. In addition, it should be a multistage process, because first of all it is necessary to determine the hypothetical phase composition, create the corresponding basic diffraction patterns, etc. And only then to calculate the relative content of phases according to the recommendations of the ''Formulas for content calculating'' section.
	\item However, it is shown that the proposed method, at least roughly within the limits of the specified uncertainty, allows identifying phases and determining their relative content (provided that the underlying approximations are correct: equality of absorption coefficients, absence of texture, etc).
	\item The proposed method can also be used as a metrological technique for checking the quality of standards.
\end{enumerate}

The authors are grateful to Mykhailo Ihor Fedorovych (National Technical University ''Kharkiv Polytechnic Institute'', Faculty of Physics and Technology, Department of Physics of Metals and Semiconductors) for providing reference diffraction patterns of mixtures of nantokite and eriochalcite.

The work was sponsored in the framework of the budget theme of the National Academy of Sciences of Ukraine
(No. 0120U103480).

\end{document}